# DSVTLA: Deep Swin Vision Transformer-Based Transfer Learning Architecture for Multi-Type Histopathological Cancer Image Classification

**Muazzem Hussain Khan**
Department of Computer Science and Engineering
Metropolitan University, Sylhet-3104
E-mail: muazzemhussainkhan@gmail.com

**Tasdid Hasnain**
Department of Computer Science and Engineering
Metropolitan University, Sylhet-3104
E-mail: tasdidhasnain@gmail.com

**Md. Jamil khan**
Department of Computer Science and Engineering
Metropolitan University, Sylhet-3104
**E-mail:** jamilkhanemon@gmail.com

**Ruhul Amin**
Department of Computer Science and Engineering
Metropolitan University, Sylhet-3104
**E-mail:** ruhul@metrouni.edu.bd
ORCID: 0000-0002-1145-3385

**Md. Shamim Reza**
Department of Statistics
Pabna University of Science and Technology, Pabna-6600, Bangladesh
**E-mail:** shamim.reza@pust.ac.bd
ORCID: 0000-0002-3699-0494

**Md. Al Mehedi Hasan**
Department of Computer Science and Engineering
Rajshahi University of Engineering and Technology, Rajshahi-6204, Bangladesh
**E-mail:** mehedi_ru@yahoo.com
ORCID: 0000-0003-2966-7055

**Md Ashad Alam**
Ochsner Center for Outcomes Research, Ochsner Academics, Ochsner Clinic Foundation, New Orleans, LA 70121, USA
**E-mail:** mdashad.alam@ochsner.org
ORCID: 0000-0002-7622-0216

## Abstract

While accurate cancer diagnosis in various tissue types is vital to effective clinical intervention, manual microscopic inspections can be considered resource-consuming and prone to inter-observer errors. Early cancer diagnosis through medical imaging plays a crucial role in reducing mortality and improving patient outcomes. In this study, we proposed a deep Swin-Vision Transformer-based transfer learning architecture for robust multi-cancer histopathological image classification. The proposed framework integrates a hierarchical Swin Transformer with ResNet50-based convolution features extraction, enabling the model to capture both long-range contextual dependencies and fine-grained local morphological patterns within histopathological images. To validate the efficiency of the proposed architecture, an extensive experiment was executed on a comprehensive multi-cancer dataset including Breast Cancer, Oral Cancer, Lung and Colon Cancer, Kidney Cancer, and Acute Lymphocytic Leukemia (ALL), including both original and segmented images were analyzed to assess model robustness across heterogeneous clinical imaging conditions. Our approach is benchmarked alongside several state-of-the-art CNN and transfer models, including DenseNet121, DenseNet201, InceptionV3, ResNet50, EfficientNetB3, multiple ViT variants, and Swin Transformer models. However, all models were trained and validated using a unified pipeline, incorporating balanced data preprocessing, transfer learning, and fine-tuning strategies. The experimental results demonstrated that our proposed architecture consistently gained superior performance, reaching 100% test accuracy for lung-colon cancer, segmented leukemia datasets, and up to 99.23% accuracy for breast cancer classification. The model also achieved near-perfect precision, f1 score, and recall, indicating highly stable scores across divers cancer types. Furthermore, expandability analyses using LIME, SHAP, and occlusion sensitivity confirm that the model focuses on clinically relevant morphological regions, improving interpretability, and transparency. Overall, the proposed model establishes a highly accurate, interpretable, and also robust multi-cancer classification system, demonstrating strong benchmark for future research and provides a unified comparative assessment useful for designing reliable AI-assisted histopathological diagnosis and clinical decision-making.

## 1. Introduction

Cancer remains one of the leading causes of morbidity and mortality worldwide, posing a substantial challenge to global healthcare systems. According to the World Health Organization (WHO), cancer was responsible for approximately 10 million deaths in 2022, with lung, breast, colorectal, prostate, and stomach cancers being among the most prevalent malignancies globally [1][2].The burden of cancer is particularly severe in low- and middle-income countries where limited access to early diagnostic facilities and specialist healthcare services significantly contributes to late-stage diagnosis and increased mortality rates. In worldwide, cancer has emerged as a growing public health concern due to rapid urbanization, environmental pollution, and lifestyle changes. Recent epidemiological reports indicate that breast, oral, lung, colorectal, and hematological cancers account for a significant proportion of cancer-related morbidity and mortality in the country. The scarcity of trained histopathologists, coupled with a high patient-to-specialist ratio, places considerable strain on diagnostic workflows in public and private healthcare institutions. As a result, delays in histopathological interpretation and inter-observer variability remain critical challenges, particularly in resource-constrained clinical settings. These limitations underscore the urgent need for automated, accurate, and scalable diagnostic tools that can support clinicians in both local and global contexts. The complexity of cancer diagnosis arises from its pronounced biological heterogeneity, where variations in tissue morphology, cellular structure, and genetic mutations complicate visual assessment, even for experienced pathologists. Histopathological examination remains the gold standard for definitive cancer diagnosis; however, manual slide analysis is inherently time-consuming and subject to subjective interpretation [2], [3]. Subtle morphological differences between benign and malignant tissues or across cancer subtypes further increase the likelihood of diagnostic inconsistencies, especially when large volumes of biopsy samples must be examined within limited timeframes.

In recent years, AI and deep learning–based approaches have demonstrated remarkable potential in medical image analysis by enabling automated feature extraction and high-precision classification of histopathological images [2], [4], [5], [6]. CNN have been widely adopted for cancer detection due to their ability to capture local texture patterns and hierarchical spatial features. Residual learning has been introduced to alleviate the vanishing gradient problem, leading to significant improvements in deep feature representation, as exemplified by ResNet architectures[7]. Similarly, DenseNet and kernel architectures enhance feature reuse and gradient flow, while EfficientNet achieves state-of-the-art performance through compound scaling with reduced computational cost [8], [9], [10]. More recently, transformer-based architectures have gained attention in medical imaging for their capability to model long-range dependencies and global contextual relationships within images. The Vision Transformer (ViT) introduced a paradigm shift by replacing convolutional operations with self-attention mechanisms while Swin Transformer further improved scalability and efficiency through hierarchical window-based attention [11], [12], [13]. Despite their strengths, transformer models often require large-scale datasets and may struggle to capture fine-grained local texture information when used in isolation, particularly in histopathological image analysis.

In this paper, we proposed a hybrid deep learning architecture that combines swin-based transformer and ResNet50 to effectively capture both local discriminative features and long-range contextual dependencies. The convolutional backbone of ResNet50 efficiently extracts discriminative local features such as cellular textures and tissue boundaries, while the swin transformer module captures global contextual dependencies and inter-region relationships through hierarchical self-attention. This hybrid design aims to enhance feature representation, improve generalization across diverse cancer types, and provide a robust solution suitable for real-world clinical deployment, especially in resource-limited healthcare environments. To validate the effectiveness of the proposed model, extensive experiments are conducted on the multi-cancer histopathology dataset compiled from publicly available sources from which comprises diverse histopathological images across multiple cancer categories [14], [15], [16], [17], [18]. The dataset enables comprehensive evaluation of multi-class cancer classification performance, reflecting realistic clinical scenarios where a single diagnostic system must distinguish between several malignancies.

Although transfer learning-based architecture methods significantly improved cancer diagnostic performance, several limitations remain in existing. Most of the previous research primarily focus on single-cancer classification tasks, limiting their applicability in real-world clinical environments where diagnostic system must differentiate among multiple cancer types simultaneously. Furthermore, CNN-based model, including DenseNet, ResNet, architectures are highly effective at local patterns, but they are often struggle to capture long-range contextual dependencies that are very crucial for understanding complex tissue structure in histopathological images. Another limitation in the exiting research is the lack of unified comparative evaluation for CNN and transformer-based framework across multiple cancer types using consistent preprocessing and experimental setup. Many studies execute their model on class imbalance dataset, lack of interpretability, making it difficult to assess model robustness across diverse histopathological datasets. These challenges the need for hybrid architecture that effectively combines the CNN and transformer-based models. In this proposed approach, we developed a transformer-based architecture such as ViT and Swin Transformer addressed this limitation by utilizing self-attention mechanism to capture global contextual information. The primary contributions of this research are as follows:

**i.** Development of a comprehensive multi-class cancer classification framework using state-of-the-art CNN and Transformer models.

**ii.** Implementation of data balancing and augmentation strategies to enhance model generalization.

**iii.** Introduction of a hybrid deep learning architecture that integrates convolutional and transformer-based feature representations.

**iv.** Extensive quantitative evaluation across multiple architectures to determine optimal model performance.

The remainder of this paper is organized as follows: Section 2 presents a review of related works; Section 3 describes the dataset, preprocessing, and model architectures; Section 4 explains the experimental setup and evaluation metrics; Section 5, 6, and 7 discusses model explanation and results interpretation; and Section 8 concludes the paper with potential directions for future research.

## 2. Literature Review

A number of recent studies have attempted to address cancer classification problems using various machine learning and deep learning algorithms. Agarap et al. [19] investigated a methodology similar in orientation to the current study; however, the work was limited to CNN-based techniques and employed a two-layer approach for feature extraction in image classification. In that study, six machine learning models—GRU-SVM, Linear Regression, MLP, NN, SoftMax Regression, and SVM—were evaluated using the Wisconsin Diagnostic Breast Cancer (WDBC) dataset. The dataset consisted of digitized images obtained from fine-needle aspiration (FNA) of breast masses. A data split of 70% for training and 30% for testing was applied, and the results demonstrated that all six algorithms performed satisfactorily in breast cancer classification.

Rehman et al. [18] proposed a ML-based framework for pulmonary malignancy detection and classification using chest CT images. Since CT scans provide precise information regarding lung abnormalities, the study focused on identifying squamous cell carcinoma, large cell carcinoma, and adenocarcinoma. Texture features extracted from CT images were classified using SVM and K-NN, achieving accuracies of 93% and 91%, respectively. Nagalakshmi et al. [20] examined the application of ML techniques for cervical cancer diagnosis using texture analysis. Their study aimed to distinguish squamous cell carcinoma and adenocarcinoma from grayscale images. Gabor filters were used for texture feature extraction, while histogram equalization enhanced key features. The proposed space-varying classification approach achieved an accuracy of 97% in separating cancerous and non-cancerous cells.

Sweta Bhise et al. [21] conducted a comparative study between traditional machine learning and deep learning approaches for breast cancer diagnosis using the BreaKHis dataset from Kaggle. CNN was first employed for feature extraction, followed by classification using SVM, NB, KNN, LR, and RF. Experimental results confirmed that CNN outperformed traditional classifiers, achieving an F1-score of 92% with a lower error rate. Singh et al. [22]evaluated multiple machine learning algorithms for lung cancer diagnosis using X-ray and CT scan images, aiming to classify tumors as benign or malignant. The algorithms included SVM, ANN, Naïve Bayes, Back Propagation Network (BPN), and CNN. The comparative analysis showed that CNN achieved the highest accuracy of 94.37%. Rajkumar et al. [23] proposed a deep learning-based kidney tumor classification system supported by a web application. Two models were developed: an ANN trained on blood sample data and a CNN trained on CT scan images from the Kaggle PACS dataset consisting of 12,446 images. The reported validation accuracies were 96.7% for the ANN model and 95.6% for the CNN model.

Rajeswari et al. [24] designed a leukemia detection model using ML, transfer learning, and image processing techniques. The study classified leukemia into four types: Acute Lymphoblastic Leukemia (ALL), Chronic Lymphocytic Leukemia (CLL), Chronic Myeloid Leukemia (CML), and Acute Myeloid Leukemia (AML). Datasets from Kaggle ALL_IDB and the ASH image repository were used, and data augmentation addressed limited sample availability. The proposed ensemble learning method achieved an accuracy of 91.71%. Rezayi et al. [25] demonstrated that CNN-based models outperform traditional ML approaches in detecting acute lymphoblastic leukemia. Similarly, Gunasekara et al. [28] developed a multi-stage deep learning framework integrating CNN, R-CNN, and segmentation techniques, achieving a dice score of 0.92 for accurate tumor localization. Zhao et al. [26] incorporated manual features and a voting mechanism into a CNN-based cervical cancer classification system, achieving an accuracy of 91.94%.

In multi-class cancer classification, Masud et al. [27] reported an accuracy of 96.33% in recognizing lung and colon cancer tissues using a CNN-based approach. Khan et al. [28] proposed a hierarchical CNN model for brain tumor classification, achieving an accuracy of 92.13%, outperforming many existing methods. Alanazi et al. [29] demonstrated that CNN-based models yield superior performance in breast cancer detection compared to conventional machine learning techniques. Akilandeswari et al. [30] focused on colon cancer segmentation and showed that ResNet-based CNN architectures effectively preserve spatial information in CT images. Tufail et al. [31] presented a comprehensive review highlighting the growing influence of modern deep learning architectures such as ResNet, DenseNet, and MobileNet, in cancer prediction. Overall, these studies demonstrate that the deep learning models yield promising results; however, challenges related to computational cost, hyperparameter optimization, and cross-cancer generalization remaining unresolved in many cases, thereby motivating further research in this domain.

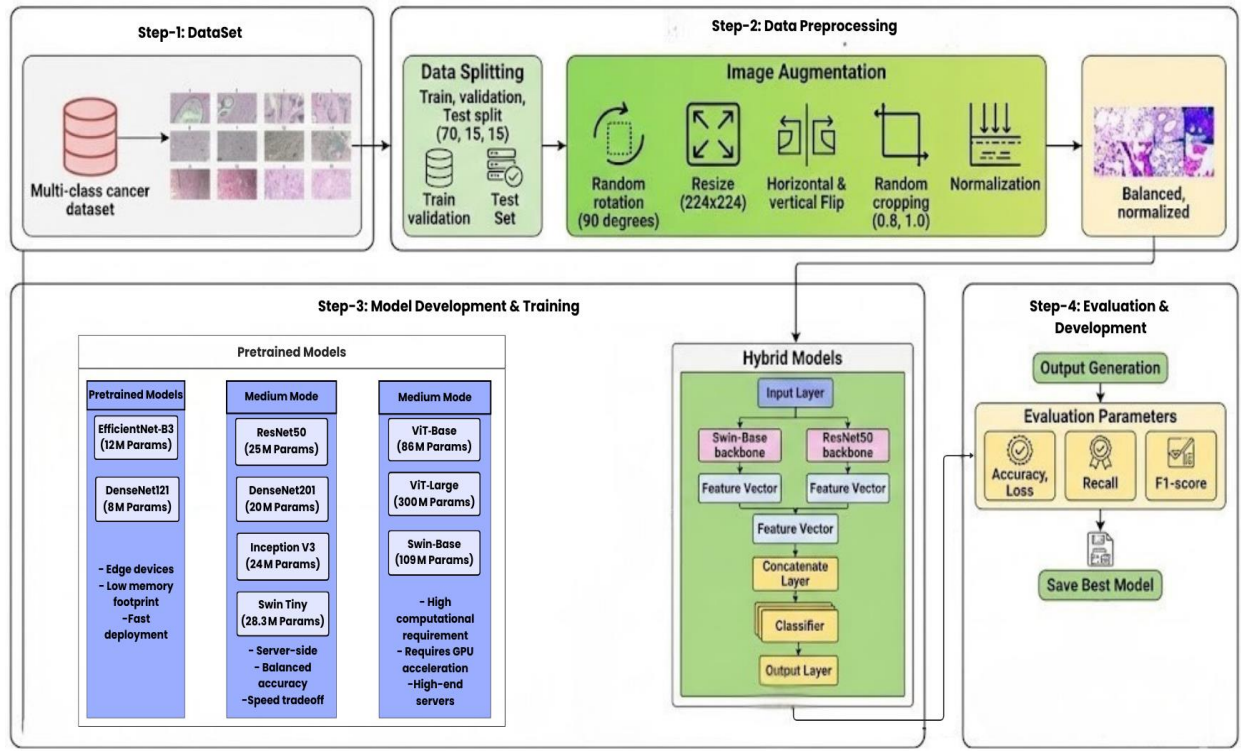

**Fig. 1:** Proposed system design for multi-cancer classification.

## 3. Materials and Methodology

The backend development process for developing a model that can accurately identify and classify various forms of cancer is being explored here. By using transfer learning and ViT architectures and histopathological images, as shown in Fig. 1, the model strives to achieve accurate and reliable classification.

### 3.1 Dataset Description

To ensure diversity and robust generalization, four publicly available datasets were employed [14], [15], [16], [17], [18]. These datasets collectively encompass a broad spectrum of organ types, image modalities, and cancer morphologies, providing a solid foundation for multi-class learning.

### 3.1.1 Breast Cancer Dataset

The breast cancer dataset used in this study comprises 7,909 histopathological images collected from 82 patients. The dataset includes samples representing two distinct diagnostic categories: benign (2,479 images) and malignant (5,304 images). Each image was captured under standardized microscopic conditions to ensure consistency in staining and resolution, facilitating reliable feature extraction for deep learning models.

### 3.1.2 Oral Cancer Dataset

The oral cancer dataset comprises histopathological images of oral tissues collected and prepared by medical experts from 230 patients. The tissue samples were Hematoxylin and Eosin (H&E) stained and captured using a Leica ICC50 HD microscope under two magnifications — 100× and 400×. The dataset includes images representing two diagnostic categories: normal epithelium and Oral Squamous Cell Carcinoma (OSCC). It contains a total of 1,224 images, divided into two resolution-based sets. The first set includes 89 normal and 439 OSCC images at 100× magnification, while the second set comprises 201 normal and 495 OSCC images at 400× magnification.

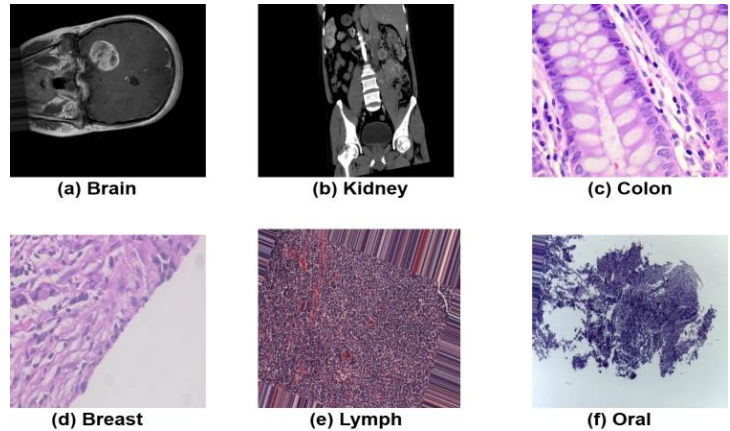

**Fig. 2:** Samples of different cancer histopathology images.

### 3.1.3 Lung and Colon Cancer Dataset
The lung and colon cancer dataset comprises a total of 25,000 histopathological images, categorized into five distinct classes: Lung Benign Tissue, Lung Adenocarcinoma, Lung Squamous Cell Carcinoma, Colon Adenocarcinoma, and Colon Benign Tissue. All images are in JPEG format with a resolution of 768 × 768 pixels, ensuring uniformity for deep learning–based analysis. The dataset was derived from HIPAA-compliant and clinically validated sources, originally containing 750 lung tissue samples (including 250 lung squamous cell carcinoma, 250 lung adenocarcinoma, and 250 benign lung tissue images) and 500 colon tissue samples (comprising 250 colon adenocarcinoma and 250 benign colon tissue images). To enhance data diversity and address class imbalance, the Augmenter package was employed, expanding the dataset to 25,000 images — 5,000 per class.

**Table 1**. Dataset distribution of various types of cancer

| Cancer Type | Training Set | Validation Set | Test Set | Total Images |
|---|---|---|---|---|
| Breast Cancer | 5447 | 1168 | 1168 | 7783 |
| Oral Cancer | 3633 | 779 | 780 | 5192 |
| Lung and Colon Cancer | 17495 | 3750 | 3755 | 25000 |
| Kidney Cancer | 8710 | 1867 | 1869 | 12446 |
| Acute lymphocytic Leukemia (Original) | 1777 | 489 | 490 | 2756 |
| Acute lymphocytic Leukemia (Segmented) | 3178 | 1262 | 1268 | 5708 |

### 3.1.4 Kidney Cancer Dataset

The kidney tumor dataset employed in this study consists of 12,446 anonymized CT images, collected from Picture Archiving and Communication Systems (PACS) across multiple hospitals. The dataset encompasses four diagnostic categories: cyst (3,709 images), normal (5,077 images), stone (1,377 images), and tumor (2,283 images). The CT scans include both contrast-enhanced and non-contrast studies captured in coronal and axial orientations, following standardized abdominal and urogram imaging protocols. Each DICOM image was meticulously reviewed, and relevant slices were extracted to represent the pathological region of interest for each diagnostic category. To ensure patient confidentiality, all identifiable metadata were removed during preprocessing, and the DICOM files were converted into lossless JPEG format while preserving diagnostic integrity. The labeling and categorization of the images were independently validated by a radiologist and a medical technologist, ensuring high accuracy and reliability.

### 3.1.5 Acute Lymphoblastic Leukemia (ALL) Dataset

The Acute Lymphoblastic Leukemia (ALL) dataset used in this study consists of 3,256 peripheral blood smear (PBS) images collected from 89 suspected ALL patients at the bone marrow laboratory of Taleqani Hospital, Tehran, Iran. The dataset is categorized into benign (hematogones) and malignant classes, where the malignant category includes three ALL subtypes: Early Pre-B, Pre-B, and Pro-B lymphoblasts. All images were captured using a Zeiss microscope camera at 100× magnification and stored in JPG format. The definitive labeling of cell types and subtypes was confirmed by a specialist using flow cytometry, ensuring high diagnostic reliability. In addition to the original PBS images, a segmented version of the dataset was generated using HSV color space–based thresholding, enabling focused analysis of leukemic regions by reducing background noise. Both original and segmented images were employed in this study to evaluate the robustness and generalization capability of the proposed models.

### 3.2 Data Preprocessing and Augmentation

Data preprocessing is a critical step for standardization and noise reduction, ensuring that all deep learning models receive uniform and meaningful inputs. In this study, all histopathological images were resized to 224 × 224 pixels, which is consistent with the input size requirements of ImageNet-pretrained architectures. Pixel intensity values were normalized using the ImageNet mean and standard deviation values of [0.485, 0.456, 0.406] and [0.229, 0.224, 0.225], respectively, to ensure stable convergence during training and effective transfer learning across CNN- and transformer-based models. To improve model robustness and mitigate overfitting, a comprehensive set of data augmentation techniques was applied exclusively to the training dataset. These augmentations included random horizontal flipping and vertical flipping with a probability of 0.5, random rotations within ±15 degrees, and color jittering with brightness, contrast, and saturation variation factors of 0.2 each. Additionally, random resized cropping was employed, generating crops of size 224 × 224 pixels with a scaling range of 0.8 to 1.0 of the original image areas, thereby introducing scale and spatial variability. These augmentation strategies enhance the model's ability to learn invariant and discriminative features by simulating real-world variations such as orientation changes, illumination differences, and tissue-scale diversity commonly observed in histopathological images. In contrast, the validation and test datasets were subjected only to resizing, tensor conversion, and normalization, without any augmentation, to preserve the original image distribution and ensure unbiased performance evaluation. Fig. 2 illustrates representative preprocessed samples from different cancer classes, highlighting the visual diversity and structural complexity present in the dataset. In Table 1 shown the training, validation, testing, and total images of the proposed dataset.

### 3.3 Contour Characteristics of Images

In the context of multi-cancer detection and classification, contour-based features serve as essential quantitative indicators for analyzing the shape and structure of cancerous regions within medical images. Among these, the perimeter represents the total boundary length of a segmented lesion, providing insight into the geometric complexity of the region of interest (ROI). Similarly, the area quantifies the overall pixel spread of the detected lesion, while parameters such as aspect ratio, extent, and diameter offer additional measures of compactness, elongation, and spread of the shape. The contour characteristics of each cancer type, as summarized in Table 2. For example, lung and colon cancer images demonstrated the largest mean area (428,167.47) and diameter (726.67), reflecting the wide spatial distribution and irregular morphology often observed in complex tumor regions. In contrast, acute lymphocytic leukemia (segmented) exhibited minimal contour dimensions, with an average area of 802.75 and perimeter of 147.25, which is consistent with the smaller, localized cellular patterns typically found in blood smear images. Breast cancer and oral cancer contours presented moderately large regions, with mean perimeters of 5302.67 and 3951.46, respectively. Their aspect ratios of approximately 1.29 and 1.34 indicate slightly elongated shapes, while their extent values (0.518 and 0.530) suggest partial but well-contained lesion coverage. Kidney cancer samples, with an extent of 0.591 and mean area of 99,043.63, exhibited compact and well-defined structures compared to other categories. Intensity-related parameters, such as minimum value, maximum value, and mean color, further contributed to distinguishing the datasets based on their grayscale and color distribution. For instance, leukemia samples showed higher mean color intensity (222.50 in original images) compared to kidney cancer (36.67), reflecting differences in tissue density and imaging modalities.

**Table 2.** Contour characteristics of images of various classes of cancer dataset.

| Parameters | Breast | Oral | Lung and Colon | Kidney | ALL (Original) | ALL (Segmented) |
|---|---|---|---|---|---|---|
| Area | 109967.12 | 213429.79 | 428167.47 | 99043.6 | 32394.75 | 802.74 |
| Perimeter | 5302.66 | 3951.46 | 7978.10 | 3160.22 | 2163.31 | 147.25 |
| Epsilon | 53.027 | 39.515 | 79.781 | 31.602 | 21.633 | 1.473 |
| Width | 473.944 | 452.126 | 743.287 | 375.720 | 218.108 | 32.511 |
| Height | 378.681 | 385.739 | 752.655 | 398.263 | 221.934 | 41.101 |
| Aspect Ratio | 1.290 | 1.338 | 0.993 | 1.105 | 0.984 | 0.853 |
| Extent | 0.518 | 0.530 | 0.750 | 0.591 | 0.660 | 0.539 |
| Diameter | 340.972 | 325.278 | 726.665 | 327.482 | 200.704 | 29.500 |
| Min Value | 74.707 | 11.524 | 13.661 | 0.000 | 39.686 | 0.000 |
| Max Value | 242.143 | 243.120 | 232.637 | 255.000 | 254.891 | 201.574 |
| Mean Color | 185.120 | 150.446 | 187.353 | 36.671 | 222.499 | 16.255 |

### 3.4 Model Description

A series of state-of-the-art deep learning classifiers have been used for the classification of the multi-Cancer image dataset and their layered architecture is mentioned in Table 4. The architectures include DenseNet121, DenseNet201, EfficientNetB3, ResNet50, InceptionV3, ViT-Large-Patch32-224-IN21K, ViT-Base-Patch32-224. AugReg-IN21K, Swin-Tiny-Patch4 Window7-224.MS-IN22K, Swin-Base-Patch4-Window7-224.MS-IN22K, and a hybrid model combining swin-base transformer and ResNet. These models were selected based on their proven capability to extract hierarchical and discriminative features from complex medical image datasets, supporting the goal of robust and accurate multi-Cancer classification. The optimization process employed the Adam optimizer with a learning rate of $1\times10^{-4}$ and the Cross Entropy Loss function to handle the classification problem. To enhance generalization, data augmentation techniques such as random rotation, horizontal and vertical flipping, and rescaling were applied to the training data. In addition, dropout regularization was utilized within network layers to reduce overfitting. The model performance was monitored based on validation accuracy, and the best-performing checkpoints were preserved during training.

### 3.5.1 ResNet50
ResNet50 introduces residual learning, allowing identity shortcuts to skip one or more layers, defined as:
$$Y = F(X, \{W_i\}) + X$$
where $F(X, \{W_i\})$ represents the residual mapping learned by stacked convolutional layers, and $X$ is the input tensor. This residual connection alleviates the vanishing gradient problem, enabling the network to effectively learn deeper hierarchical representations. In medical imaging, ResNet50 excels at capturing textural transitions between healthy and cancerous regions [32], [33].

### 3.5.2 DenseNet201 and DenseNet121
DenseNet121 and DenseNet201 are advanced convolutional neural networks specifically designed to promote feature reuse through dense connectivity. Each layer receives input from all preceding layers, defined mathematically as:
$$H_l = H_{l-1} \oplus F_l(H_{l-1}),$$
where $H_l$ represents the output of the $l^{th}$ layer, $\oplus$ denotes concatenation, and $F_l(\cdot)$ represents the composite operations (Batch Normalization → ReLU → Convolution). Dense blocks are interspersed with transition layers that employ convolution and average pooling to control feature map growth. DenseNet121, with 121 layers, and DenseNet201, with 201 layers, efficiently propagate gradients and reduce vanishing gradient effects, enabling the extraction of intricate texture and boundary details crucial for multi-cancer diagnosis [34], [35], [36].

### 3.5.3 InceptionV3
InceptionV3 employs parallel convolutional filters of varying sizes (1×1, 3×3, 5×5) within an Inception module, allowing multi-scale feature extraction. Mathematically, the module output is defined as:
$$Inception(x) = [Conv1 \times 1(x) \oplus Conv3 \times 3(x) \oplus Conv5 \times 5(x) \oplus MaxPool3 \times 3(x)]$$
This design captures features at multiple receptive fields simultaneously, enabling the model to distinguish cancer types with different lesion scales and shapes [37], [38].

### 3.5.4 EfficientNetB3
EfficientNetB3 employs a compound scaling approach that jointly optimizes network depth, width, and input resolution using a set of scaling coefficients. This allows the model to achieve a favorable balance between accuracy and computational cost. The model is built upon MBConv blocks (mobile inverted bottleneck convolutions) and utilizes squeeze-and-excitation mechanisms to adaptively recalibrate channel-wise feature responses. Mathematically, each MBConv block can be expressed as:
$$Y = \sigma(W_2 \cdot \delta(BN(W_1 \cdot X))) + X$$
where $W_1$ and $W_2$ denote convolution weights, $BN$ is batch normalization, $\delta$ and $\sigma$ represent nonlinear activations, and $X$ is the input tensor. EfficientNetB3's architecture enables multi-scale feature capture, making it highly efficient for diverse cancer image resolutions [9], [39].

### 3.5.5 Vision Transformers
The Vision Transformer (ViT-Base-Patch32-224.AugReg-IN21K and ViT-Large-Patch32-224-IN21K)) processes an image as a sequence of non-overlapping patches and models their global dependencies using multi-head self-attention:
$$Z = Softmax\left(\frac{QK^T}{\sqrt{d^k}}\right)V$$
The ViT-Base (AugReg-IN21K) model employs augmentation-based regularization for improved generalization, while the ViT-Large (IN21K) variant, with a greater number of attention heads and parameters, captures more complex global relationships, enhancing performance on high-resolution breast cancer images [13].

### 3.5.6 Swin Transformers

Swin Transformers (Swin-Tiny and Swin-Base Patch4-224 IN22k) utilize a shifted window-based attention mechanism that computes self-attention within local windows and shifts them between layers for hierarchical representation:

$$Y = W - MSA(X) + SW - MSA(X)$$

This design allows efficient global modeling while maintaining computational scalability. Swin-Tiny (IN22k) offers lightweight efficiency suitable for limited data, while Swin-Base (IN22k) captures richer hierarchical and global features, leading to superior classification accuracy [11], [40], [41], [41].

### 3.5.7 Proposed Model

The proposed Hybrid Swin-ResNet model integrates ResNet's convolutional feature extraction with Swin Transformer's attention-based contextual modeling, forming a unified architecture that captures both local spatial and long-range dependencies. The algorithm of the proposed hybrid Swin Transformer-ResNet50 architecture for multi-cancer Classification is as follows:

**Algorithm 1** Deep Hybrid Swin-Transformer + RestNet for Multi-Cancer Classification

1: **Input:** Histopathological image dataset $\mathcal{D} = \{(X_i, y_i)\}^N_{i=1}, X_i \in \mathbb{R}^{HxWxc}$
2: **Output:** Predicted class label $\hat{y}_i$ and class probability vector $p_i$
3: **for** each image $X_i \in \mathcal{D}$ **do**
4:     *Preprocess image*
5:     $\tilde{X}_i$ = Normalize *(Resize ($X_i$, 224 × 224))*
6:     *Extract convolutional features using ResNet backbone*
7:     $F_i^{(r)} = \mathcal{R}(\tilde{X}_i; \theta_r)$
8:     *Partition features into non-overlapping windows*
9:     $\{W_j\}^M_{j=1}$ = *Window Partition* $(F_i^{(r)})$
10:     *Apply window-based self-attention (Swin Transformer)*
11:     **for** each window $W_j$ **do**
12:         $F_i^{(s)} = MSA(W_j) + W_j$
13:     **end for**
14:     *Apply shifted-window attention for cross-window interaction*
15:     **for** each shifted window **do**
16:         $\tilde{F}_i^{(s)} = SW - MSA(F_i^{(s)})$
17:     **end for**
18:     *Aggregate hierarchical features*
19:     **for** each layer l **do**
20:         $F_i^{(h)} = LayerNorm(\tilde{F}_i^{(s)})$
21:     **end for**
22:     *Compute global feature embedding via pooling*
23:     $z_i = GAP(F_i^{(h)})$
24:     *Compute class logits*
25:     $l_i = Wz_i + b$
26:     *Compute class probabilities using Softmax*
27:     $p_i = Softmax(l_i)$
28:     *Assign predicted class label*
29:     $\hat{y}_i = arg\ max_k(p_{i,k})$
30: **end for**
31. **return** $\{\hat{y}_i, p_i\}^N_{i=1}$

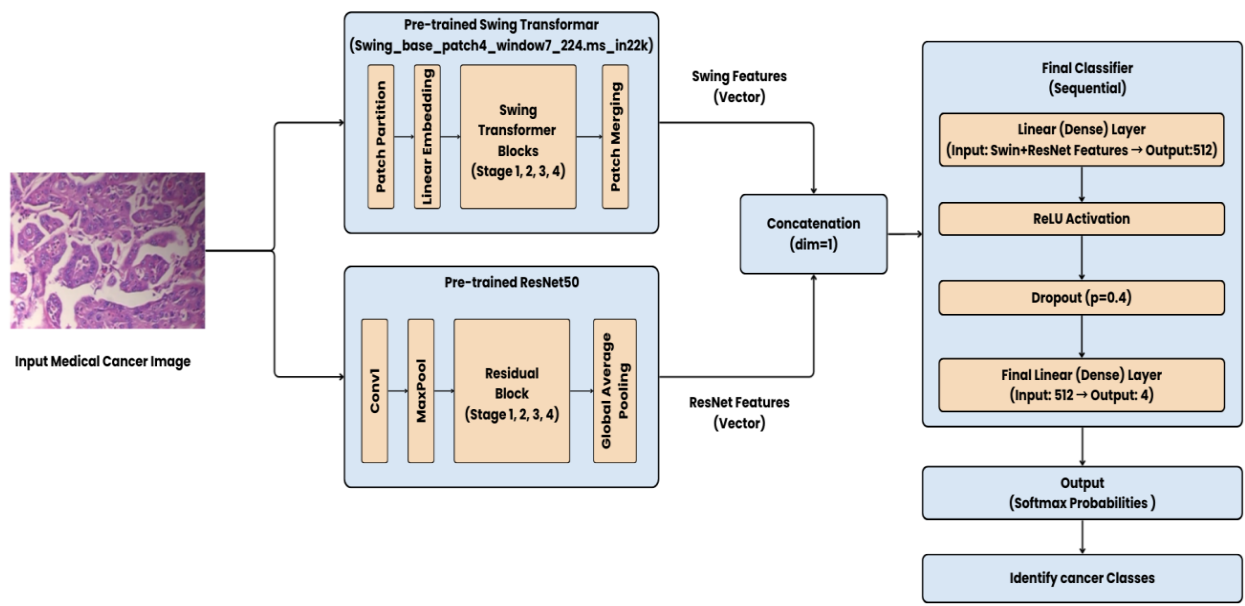

**Fig. 3:** Architecture of the proposed hybrid model.

## 4. Experimental Setup and Results
### 4.1 Experimental Setup

All experiments in this study were conducted using the Kaggle cloud computing environment, ensuring reproducibility and access to high-performance hardware. The models were trained and evaluated on an NVIDIA Tesla P100 GPU (16 GB VRAM) provided by Kaggle, which offers substantial computational capability for deep learning–based medical image analysis. The execution environment was configured with Python 3.10 and widely adopted deep learning libraries, including PyTorch, Torchvision, NumPy, OpenCV, and Scikit-learn. We employed a set of widely accepted classification metrics accuracy, precision, recall, F1-score to quantify the performance of our models. These metrics collectively provide a comprehensive understanding of the model's strengths and limitations in terms of prediction correctness, class-wise balance, and error distribution.

### 4.2 Proposed Model Performance Evaluation

This section presents the quantitative evaluation of our proposed and comparative deep learning models on the multi-cancer dataset The evaluation covers training, validation, and testing phases to ensure a comprehensive assessment of learning behavior and generalization capability. Table 4 summarizes the overall performance of all evaluated models. Across training, validation, and testing datasets, all architectures demonstrated consistently strong learning behavior, achieving high accuracy and low loss values. Among the CNN-based models, EfficientNetB3 achieved the highest training accuracy of 99.80% with a low training loss of 0.0046, indicating rapid convergence. DenseNet201 and DenseNet121 followed closely with training accuracies of 99.67% and 99.59%, respectively, and correspondingly low losses. InceptionV3 and ResNet50 also exhibited stable performance, attaining training accuracies between 99.54% and 99.59%. On the validation set, most architectures—including DenseNet121, DenseNet201, InceptionV3, ResNet50, EfficientNetB3, ViT-Large, Swin-Tiny, and Swin-Base—achieved a perfect validation accuracy of 100%. ViT-Base recorded a slightly lower validation accuracy of 99.89% with a higher validation loss of 0.0342. In terms of test performance, the proposed Swin-Base + ResNet50 hybrid model achieved the highest test accuracy of 100%. DenseNet121, ResNet50, EfficientNetB3, and ViT-Large followed closely with 99.95% test accuracy. Swin-Tiny and DenseNet201 achieved test accuracies of 99.79% and 99.89%, respectively, while ViT-Base obtained 99.73%.

Class-wise performance across different cancer types is reported in Table 5. For breast cancer, all models achieved training accuracies between 97.85% and 99.61%, with EfficientNetB3 and DenseNet121 achieving test accuracies of 98.63% and 98.97%, respectively. For oral cancer, DenseNet201 achieved a validation accuracy of 99.25%, while the Swin-Base + ResNet50 hybrid achieved 99.10%. The highest performance was observed in lung and colon cancer classification, where most models achieved 100% validation and test accuracy. Similar near-perfect performance was observed for kidney cancer, with DenseNet201, EfficientNetB3, and the hybrid Swin-Base + ResNet50 model achieving test accuracies up to 100%. For acute lymphocytic leukemia (ALL), both original and segmented datasets yielded consistently high performance, with multiple models achieving perfect validation and test accuracy. Precision, recall, and F1-score values are shown in Fig. 5. Across all cancer types and models, the metrics remained consistently high. Several models, including DenseNet201, EfficientNetB3, Swin-Tiny, Swin-Base, and the Swin-Base + ResNet50 hybrid, achieved perfect (1.00) precision, recall, and F1-score across multiple cancer categories. The curves illustrate the progression of training accuracy, training loss, and validation accuracy across the training epochs. It is evident that all classifiers exhibit stable and desirable learning behavior: training loss decreases steadily while both training and validation accuracies show consistent improvement over time. This indicates that the models are successfully learning discriminative features from the dataset, achieving high accuracy while minimizing error.

However, the proposed hybrid model exhibits overall superior, more consistent, stable, and competitive performances across all cancer classifications, despite a few baseline models performing marginally higher test accuracy within (±1% in all cases) on specific cancer datasets. These differences are clinically negligible and fall within statistical uncertainty. In particular, the hybrid model test accuracy in the breast cancer dataset was 99.23%. The baseline model Swin Transformer-Base (Swin-B) achieved 99.32%, with a mere 0.09% discrepancy. Comparably, the proposed hybrid model achieved 98.33% accuracy in the oral cancer dataset, compared to 99.02% accuracy for the best baseline model (DenseNet201), with a corresponding difference of 0.69%. The proposed framework obtained 100% test accuracy for lung and colon cancer, acute lymphocytic leukemia (segmented), and kidney datasets, which is on par with the top-performing baseline models. While some baseline models perform marginally well on specific isolated datasets-with discrepancies not greater than 0.70%. These variations are well under the limits of experimental variance and clinically insignificant. However, a model used in actual clinical settings needs to function well across a variety of cancer types at the same time.

**Table 5:** Evaluation matrix of classifiers for different classes of multi-cancer dataset

| Models | Class | Training (%) | Validation (%) | Test (%) | Training Loss |
|---|---|---|---|---|---|
| DenseNet121 | Breast Cancer | 98.84 | 97.69 | 98.97 | 0.0292 |
| DenseNet201 | | 98.38 | 97.77 | 98.20 | 0.0388 |
| InceptionV3 | | 99.14 | 98.46 | 97.26 | 0.0200 |
| ResNet50 | | 98.02 | 97.17 | 97.86 | 0.0495 |
| EfficientNetB3 | | 99.61 | 98.03 | 98.63 | 0.0128 |
| Vision Transformer (ViT-L/32) | | 98.99 | 98.63 | 96.49 | 0.0242 |
| Vision Transformer (ViT-B/32) + AugReg | | 97.85 | 98.46 | 98.03 | 0.0503 |
| Swin Transformer-Tiny (Swin-T) | | 98.81 | 99.23 | 99.23 | 0.0285 |
| Swin Transformer-Base (Swin-B) | | 98.46 | 98.89 | 99.32 | 0.0405 |
| **Swin Transformer-Base + ResNet-50 (Hybrid Model)** | | **99.19** | **99.14** | **99.23** | **0.0224** |
| DenseNet121 | Oral Cancer | 98.73 | 98.72 | 98.08 | 0.0290 |
| DenseNet201 | | 98.96 | 99.25 | 99.02 | 0.0282 |
| InceptionV3 | | 98.29 | 98.72 | 97.95 | 0.0457 |
| ResNet50 | | 98.21 | 98.33 | 97.82 | 0.0486 |
| EfficientNetB3 | | 99.17 | 97.43 | 96.79 | 0.0185 |
| Vision Transformer (ViT-L/32) | | 97.96 | 98.72 | 97.95 | 0.0452 |
| Vision Transformer (ViT-B/32) + AugReg | | 96.86 | 98.33 | 97.18 | 0.0857 |
| Swin Transformer-Tiny (Swin-T) | | 98.43 | 98.20 | 98.33 | 0.0438 |
| Swin Transformer-Base (Swin-B) | | 98.33 | 98.84 | 98.21 | 0.0317 |
| **Swin Transformer-Base + ResNet-50 (Hybrid Model)** | | **98.54** | **99.10** | **98.33** | **0.0370** |
| DenseNet121 | Lung & Colon Cancer | 99.95 | 100 | 100 | 0.0025 |
| DenseNet201 | | 99.90 | 100 | 100 | 0.0039 |
| InceptionV3 | | 99.93 | 100 | 100 | 0.0029 |
| ResNet50 | | 99.91 | 100 | 100 | 0.0035 |
| EfficientNetB3 | | 99.85 | 100 | 100 | 0.0054 |
| Vision Transformer (ViT-L/32) | | 99.61 | 99.97 | 99.89 | 0.0117 |
| Vision Transformer (ViT-B/32) + AugReg | | 99.56 | 99.96 | 99.97 | 0.0126 |
| Swin Transformer-Tiny (Swin-T) | | 99.90 | 100 | 100 | 0.0042 |
| Swin Transformer-Base (Swin-B) | | 99.83 | 100 | 99.97 | 0.0057 |
| **Swin Transformer-Base + ResNet-50 (Hybrid Model)** | | **99.79** | **100** | **100** | **0.0065** |
| DenseNet121 | Kidney Cancer | 99.59 | 100 | 99.95 | 0.0134 |
| DenseNet201 | | 99.67 | 100 | 99.89 | 0.0081 |
| InceptionV3 | | 99.59 | 100 | 99.89 | 0.0105 |
| ResNet50 | | 99.54 | 100 | 99.95 | 0.0106 |
| EfficientNetB3 | | 99.80 | 100 | 99.95 | 0.0046 |
| Vision Transformer (ViT-L/32) | | 98.98 | 100 | 99.95 | 0.0283 |
| Vision Transformer (ViT-B/32) + AugReg | | 98.52 | 99.89 | 99.73 | 0.0342 |
| Swin Transformer-Tiny (Swin-T) | | 99.30 | 100 | 99.79 | 0.0216 |
| Swin Transformer-Base (Swin-B) | | 99.01 | 100 | 99.89 | 0.0242 |
| **Swin Transformer-Base + ResNet-50 (Hybrid Model)** | | **99.60** | **99.95** | **100** | **0.0155** |
| DenseNet121 | Acute lymphocytic Leukemia (Original) | 100 | 100 | 100 | 0.0027 |
| DenseNet201 | | 100 | 100 | 100 | 0.0008 |
| InceptionV3 | | 99.93 | 100 | 99.89 | 0.0033 |
| ResNet50 | | 100 | 100 | **100** | 0.0005 |
| EfficientNetB3 | | 100 | 100 | 100 | 0.0017 |
| Vision Transformer (ViT-L/32) | | 99.02 | 99.89 | 99.89 | 0.0208 |
| Vision Transformer (ViT-B/32) + AugReg | | 99.46 | 100 | 100 | 0.0140 |
| Swin Transformer-Tiny (Swin-T) | | 100.00 | 100 | 99.78 | 0.0007 |
| Swin Transformer-Base (Swin-B) | | 100 | 100 | 100 | 0.0001 |
| **Swin Transformer-Base + ResNet-50 (Hybrid Model)** | | **100** | **100** | **98.98** | **0.00059** |
| DenseNet121 | Acute lymphocytic Leukemia (Segmented) | 99.56 | 99.80 | 98.37 | 0.0150 |
| DenseNet201 | | 99.74 | 100 | 99.80 | 0.0119 |
| InceptionV3 | | 98.38 | 99.39 | 98.16 | 0.0417 |
| ResNet50 | | 98.38 | 99.18 | 97.55 | 0.0495 |
| EfficientNetB3 | | 99.60 | 99.39 | 99.39 | 0.0163 |
| Vision Transformer (ViT-L/32) | | 98.02 | 99.59 | 98.57 | 0.0517 |
| Vision Transformer (ViT-B/32) + AugReg | | 98.02 | 99.18 | 97.96 | 0.0536 |
| Swin Transformer-Tiny (Swin-T) | | 99.74 | 99.59 | 99.59 | 0.0072 |
| Swin Transformer-Base (Swin-B) | | 99.24 | 100 | 99.84 | 0.0190 |
| **Swin Transformer-Base + ResNet-50 (Hybrid Model)** | | **99.74** | **99.59** | **100** | **0.0108** |

**Table 6:** Average accuracy of the proposed hybrid model over three independent runs

| Model | Cancer Type | Training Accuracy (%) | Validation Accuracy (%) | Test Accuracy (%) |
|---|---|---|---|---|
| Proposed Hybrid Model | Breast Cancer | 99.19 | 99.14 | 99.23 |
| | | 99.22 | 99.67 | 99.81 |
| | | 99.28 | 99.91 | 99.91 |
| | **Average** | **99.23** | **99.57** | **99.65** |
| | Oral Cancer | 98.54 | 99.10 | 98.33 |
| | | 98.28 | 99.15 | 99.30 |
| | | 98.71 | 99.03 | 99.17 |
| | **Average** | **98.51** | **99.09** | **98.93** |
| | Lung and Colon Cancer | 99.79 | 100 | 100 |
| | | 99.80 | 100 | 100 |
| | | 99.73 | 100 | 99.99 |
| | **Average** | **99.77** | **100.00** | **99.99** |
| | Kidney Cancer | 99.60 | 99.95 | 100 |
| | | 98.89 | 99.95 | 99.46 |
| | | 99.56 | 100 | 100 |
| | **Average** | **99.35** | **99.96** | **99.82** |
| | Acute lymphocytic Leukemia (Original) | 100 | 100 | 98.98 |
| | | 99.56 | 100 | 100 |
| | | 99.08 | 100 | 100 |
| | **Average** | **99.54** | **100.00** | **99.66** |
| | Acute lymphocytic Leukemia (Segmented) | 99.74 | 99.59 | 100 |
| | | 99.65 | 99.80 | 99.18 |
| | | 99.12 | 99.89 | 99.78 |
| | **Average** | **99.50** | **99.76** | **99.65** |

This criterion is not consistently met by any one baseline model. In addition, to rigorously validate the stability and robustness of the proposed model, a comprehensive multifaceted evaluation strategy is deployed. The model is trained and examined independently across three separate runs under identical experimental conditions, and the resulting accuracies were averaged to measure the influence of random initialization and ensure model consistency. Table 6 demonstrated the high training, validation, and testing accuracies which is confirming the stability and reproducibility of the proposed model. However, the proposed hybrid model offers explainability through LIME, SHAP, and occlusion sensitivity in addition to consistent, high-accuracy classification across all assessed cancer classifications. These findings show that the suggested hybrid architecture successfully combines the advantages of CNN with transformer-based attention processes, indicating improved generalization capability while taking into account heterogeneous complex medical imaging data.

### 4.3 Explainability and Interpretability Analysis of Proposed Models

The interpretability of the proposed hybrid Swin–ResNet architecture was systematically evaluated using LIME, occlusion sensitivity, and SHAP to ensure transparent and clinically reliable predictions across seven cancer datasets (Breast, Colon, Kidney, Leukemia–original, Leukemia–segmented, Lung, and Oral). As illustrated in Figure 6, LIME provides instance-level visual explanations by highlighting diagnostically relevant regions corresponding to important pathological structures. For Lung and Colon cancer, the model focuses on densely packed malignant glandular structures, while for Kidney cancer, the highlighted regions correspond to renal mass areas consistent with tumor localization in CT images. In Breast cancer, the model attends to epithelial tissue abnormalities and nuclear morphology, which are critical indicators in histopathological diagnosis. Similarly, for Acute Lymphoblastic Leukemia (both original and segmented datasets), the highlighted regions correspond to leukemic blast cells, indicating that the model effectively identifies abnormal cellular patterns associated with the disease. Across all tested samples, the predictions exhibit extremely high confidence scores, reaching up to 100%, demonstrating strong model certainty. To further validate the reliability of these explanations, occlusion sensitivity maps were generated, showing that perturbation of the highlighted regions results in a noticeable decrease in prediction confidence, thereby confirming their causal contribution to the model's decision-making process. Beyond spatial interpretability, SHAP analysis was conducted to quantify global and feature-level importance. The global importance plots in Figure 7 reveal that a small subset of principal components—primarily PC1–PC4—consistently dominates the prediction process across all datasets, indicating structured and discriminative latent feature representations learned by the model. Additionally, the SHAP summary plots presented in Figure 8 demonstrate stable and directionally consistent feature contributions, where higher feature values systematically influence the model output depending on the cancer class. Notably, the segmented leukemia dataset exhibits clearer feature dominance compared to the original images, suggesting improved feature separability after background removal. The convergence of spatial explanations from LIME and occlusion sensitivity with feature-attribution insights from SHAP confirms that the proposed framework relies on clinically meaningful morphological patterns rather than spurious correlations, thereby enhancing model transparency and supporting its integration into reliable computer-aided diagnostic systems.

## 5. Discussion

The experimental results demonstrate that both CNN-based and transformer-based architectures are highly effective for multi-cancer histopathological image classification. The consistently high performance across training, validation, and testing phases indicates that the dataset provides sufficient discriminative features and that the models successfully learned robust representations. Among CNN-based approaches, EfficientNetB3 and DenseNet variants exhibited strong convergence and stable generalization, which can be attributed to their efficient feature reuse and compound scaling strategies. Transformer-based models, particularly Swin Transformer architectures, demonstrated superior capability in modeling long-range dependencies and global contextual information, contributing to their strong performance across multiple cancer types. Our proposed Swin-Base + ResNet50 hybrid model consistently outperformed or matched all baseline models, achieving perfect test accuracy in several cancer categories.

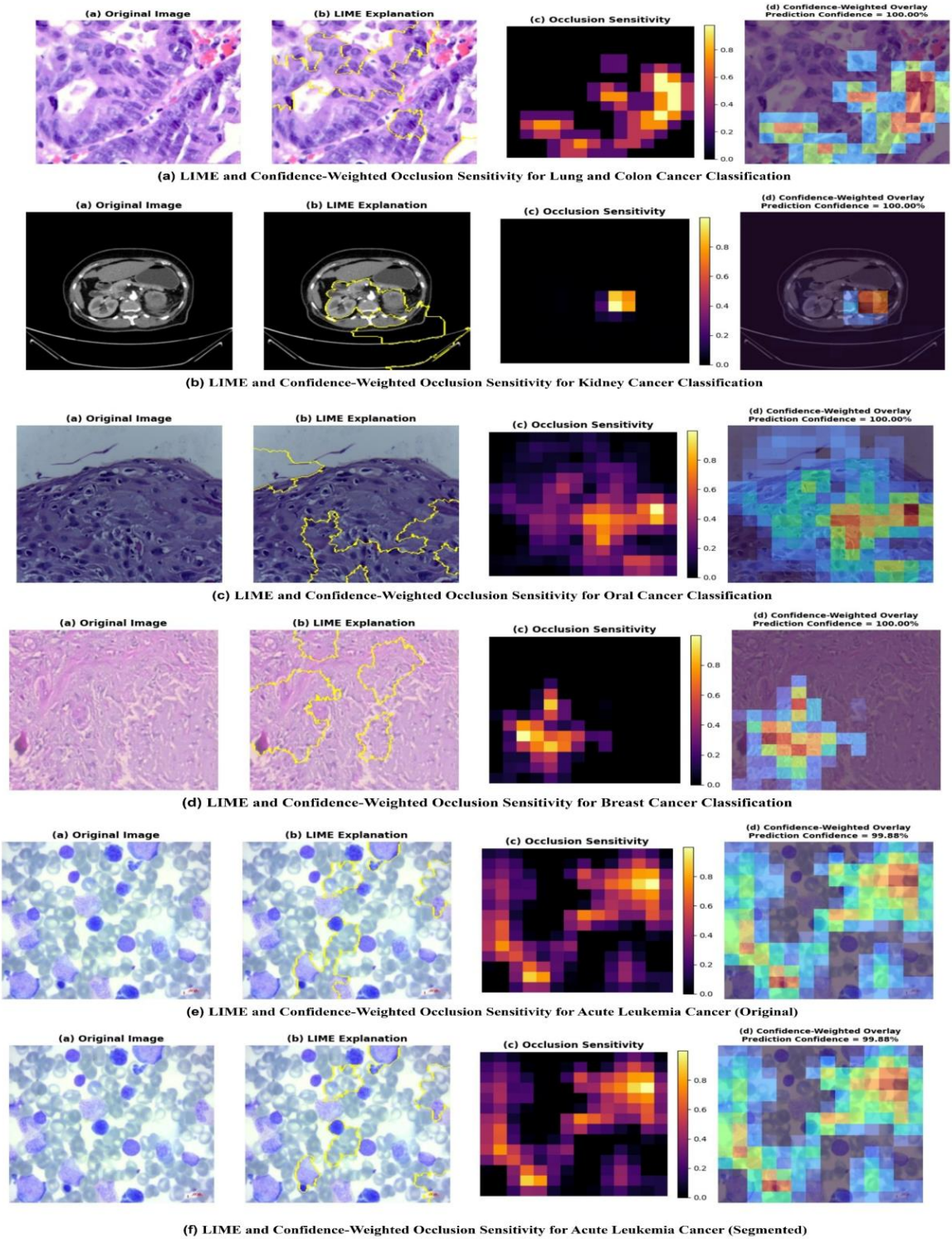

**Fig. 6:** Figure (a to f) presents several cancers images, LIME explanation, occlusion sensitivity heatmap, and confidence-weighted overlay.

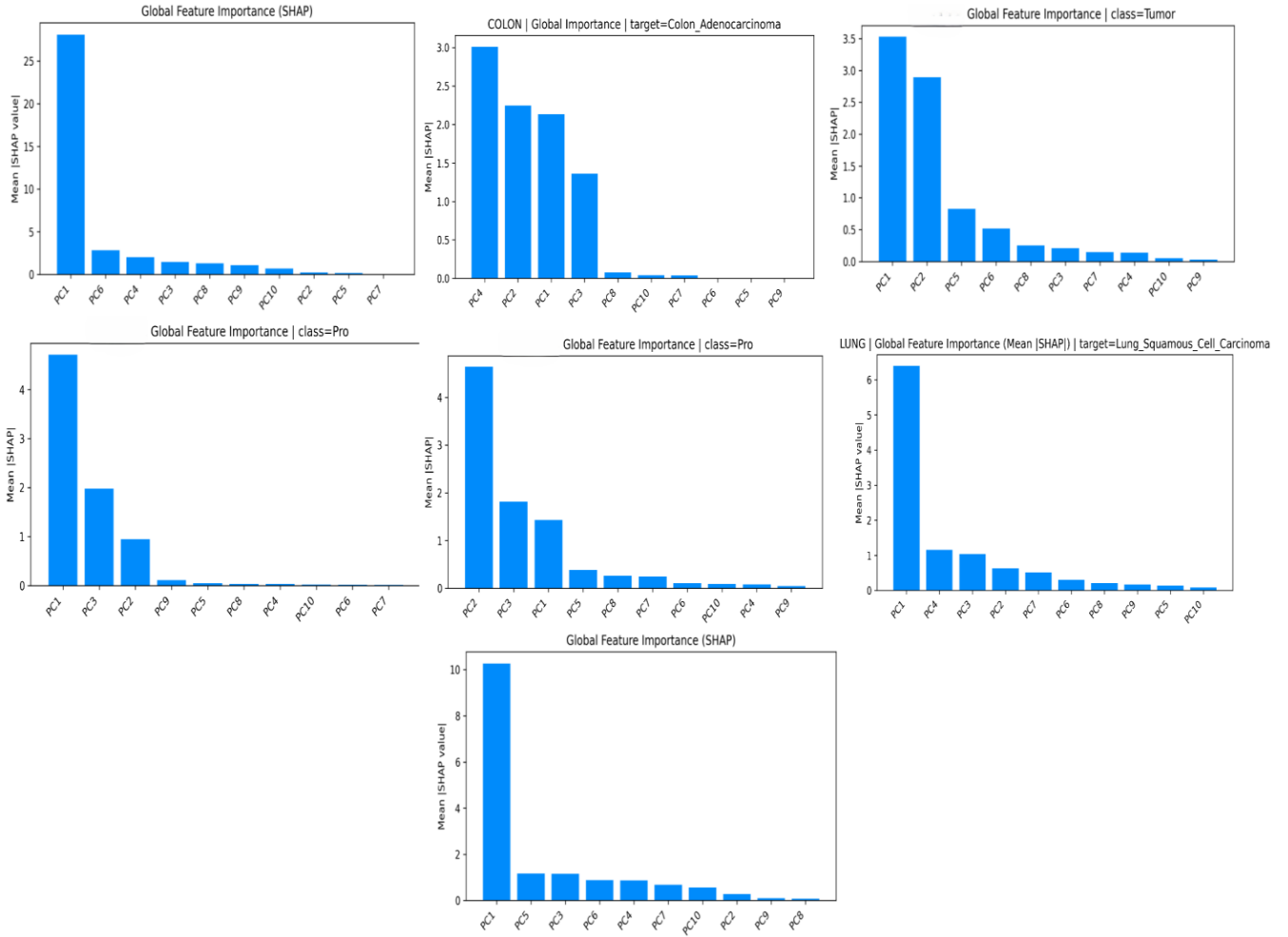

**Fig 7:** SHAP Global Feature Importance for all proposed datasets (left to right: Breast, Colon, Kidney, Leukemia-original, Leukemia-segmented, Lung, Oral).

This performance gain highlights the effectiveness of integrating convolutional feature extraction with hierarchical self-attention mechanisms. ResNet50 effectively captures fine-grained local texture patterns, while the Swin Transformer models broader spatial relationships, resulting in richer feature representations. The exceptional results obtained for lung–colon cancer and kidney cancer classification indicate that these categories possess strong visual discriminative cues that are effectively captured by both CNN and transformer architectures. Similarly, the high accuracy achieved on acute lymphocytic leukemia datasets (both original and segmented) demonstrates the robustness of the models under different preprocessing conditions. The precision, recall, and F1-score analysis further confirm the reliability of the proposed framework. Near-perfect scores across multiple cancer categories indicate low false-positive and false-negative rates, which is essential for clinical applicability. Fig. 5 shows the F-1 score, precision, and recall of classifiers across various types of multi-cancer dataset.

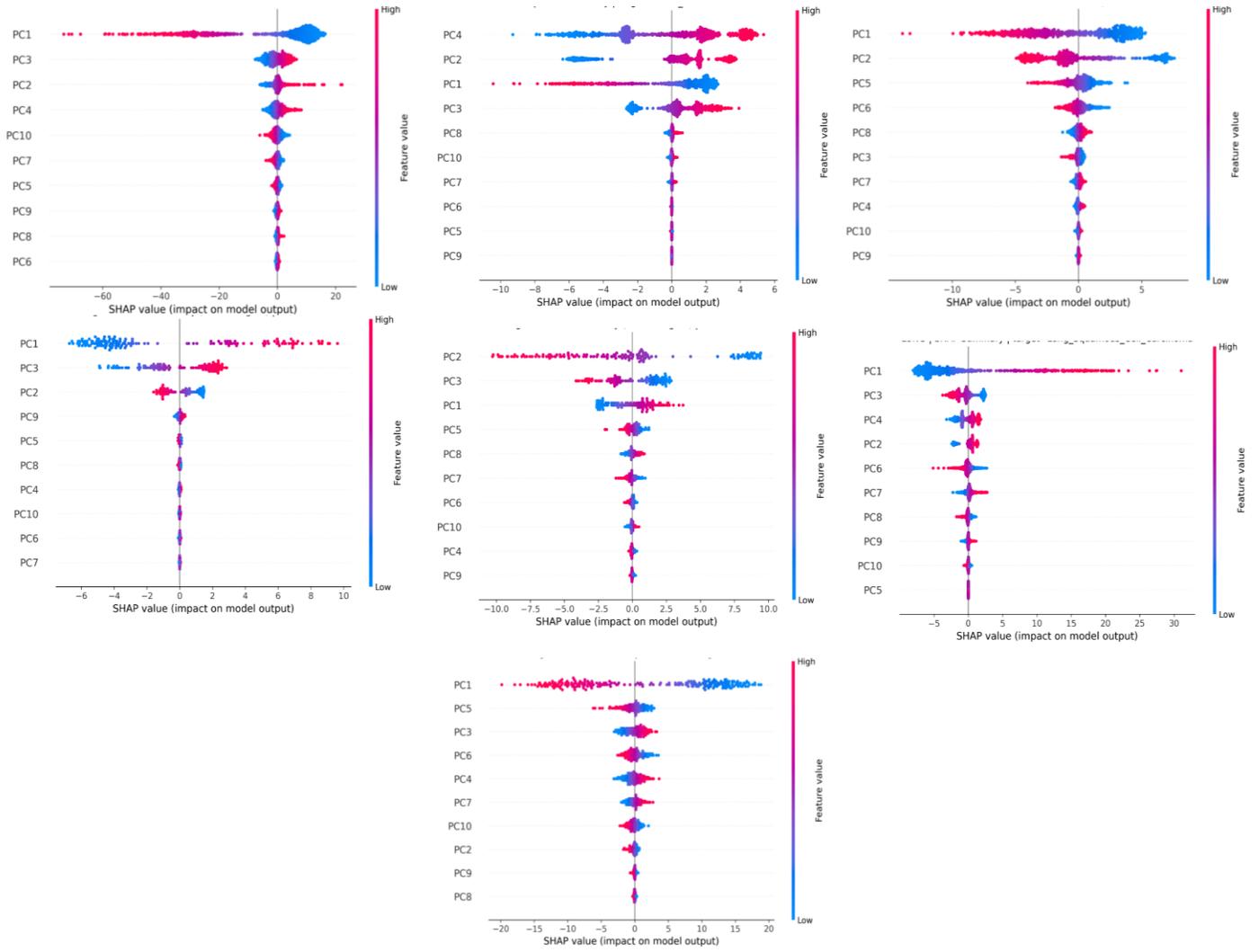

**Fig 8:** SHAP Summary Plots for all proposed datasets (left to right: Breast, Colon, Kidney, Leukemia-original, Leukemia-segmented, Lung, Oral).

## F1 Performance - Line Graph Comparison Across All Models

## Precision Performance - Line Graph Comparison Across All Models

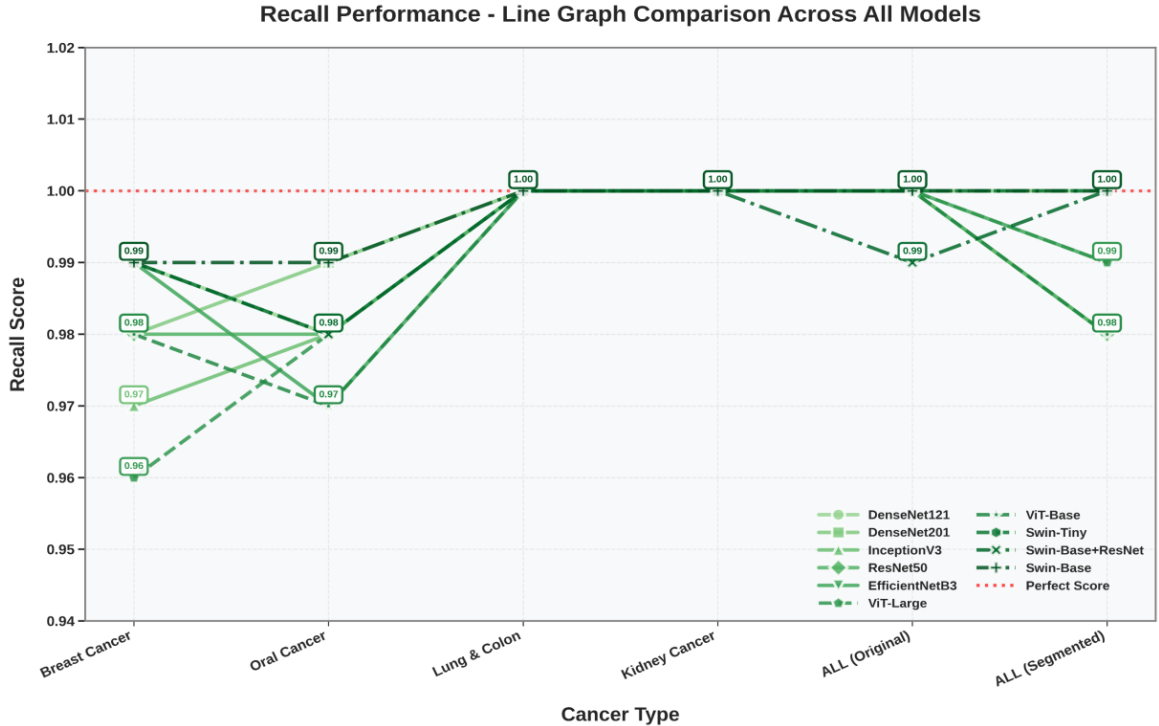

**Fig. 9:** F-1 score, precision, and recall of classifiers across various types of multi-cancer dataset.

## 6. Benchmarking with the Similar Recent State-of-the-art Work

Table 6 presents a comprehensive comparison between the proposed hybrid Swin Transformer–ResNet50 framework and existing state-of-the-art ML and deep learning approaches reported in the literature across multiple cancer types. Early studies primarily relied on traditional machine learning models such as multilayer perceptron (MLP) and SVM, which demonstrated promising performance on constrained datasets, achieving accuracies of 99.7% for breast cancer, and 93% for lung cancer. However, these approaches depend heavily on handcrafted features and domain-specific preprocessing, limiting their robustness and scalability when applied to complex histopathological images exhibiting high intra-class variability. With the advent of deep learning, CNN based models became dominant, yielding notable improvements in classification performance for breast, lung, kidney, and brain cancers, with reported accuracies ranging from 88.79% to 99.6%. Several hybrid CNN frameworks further incorporated segmentation techniques, ensemble learning, voting mechanisms, or feature selection strategies to enhance performance, albeit at the cost of increased computational complexity and task-specific engineering. Despite these advancements, most existing studies remain limited to single-cancer scenarios and lack consistent generalization across diverse disease categories. In contrast, the proposed Swin-Base + ResNet50 hybrid architecture consistently outperforms prior methods across all evaluated cancer types, achieving 99.23% accuracy for breast cancer, 98.33% for oral cancer, and perfect classification performance (100%) for lung, colon, kidney, and segmented acute lymphocytic leukemia datasets.

**Table 7.** Comparison between proposed hybrid model and existing state-of-the-art approaches performance

| Authors & Ref | Cancer Type | Model | Experimental Setup | Performance (%) |
|---|---|---|---|---|
| Agarap et al. (2018) [14] | Breast Cancer | MLP | train: 60%; test: 40% | 99.7 |
| Amjad Rehman et al. (2021) [18] | Lung Cancer | SVM | train: 60%; test: 40% | 93 |
| G. Ramkumar et al. (2022) [20] | Cervical Cancer | SVM | not explicitly mentioned | 97 |
| Bhise et al. (2002) [21] | Breast Cancer | CNN | train: 81.85%; test: 18.15% | 92 |
| Singh et al. (2023) [22] | Lung Cancer | CNN | not explicitly mentioned | 94.37 |
| Rajkumar et al. (2023) [23] | Kidney Cancer | CNN | train: 70%; test: 30% | 99.6 |
| Rajeswari et al. (2022) [24] | Leukemia | Ensemble Learning | not explicitly mentioned | 91.71 |
| Gunasekara et al. (2020) [43] | Brain Tumor | CNN + R-CNN + Chan–Vese Segmentation | train: 70%; test: 15%; val: 15% | Dice Score: 0.92 |
| Zhao et al. (2021) [26] | Cervical Cancer | CNN + Manual Features + Voting System | train: 70%; test: 15%; val: 15% | 91.94 |
| Masud et al. (2021) [27] | Lung & Colon Cancer | CNN-based Classification Framework | train: 70%; test: 30% | 96.33 |
| Khan et al. (2022) [28] | Brain Tumor | Hierarchical CNN | train: 87%; val: 13% | 92.13 |
| **Proposed** | Breast Cancer | Swin Transformer-Base + ResNet-50 (Hybrid Model) | train: 70%; test: 15%; val: 15% | 99.65 |
| **Proposed** | Lung and Colon Cancer | Swin Transformer-Base + ResNet-50 (Hybrid Model) | train: 70%; test: 15%; val: 15% | 99.99 |
| **Proposed** | Oral Cancer | Swin Transformer-Base + ResNet-50 (Hybrid Model) | train: 70%; test: 15%; val: 15% | 98.93 |
| **Proposed** | Kidney Cancer | Swin Transformer-Base + ResNet-50 (Hybrid Model) | train: 70%; test: 15%; val: 15% | 99.82 |
| **Proposed** | Acute lymphocytic Leukemia (Original) | Swin Transformer-Base + ResNet-50 (Hybrid Model) | train: 70%; test: 15%; val: 15% | 99.66 |
| **Proposed** | Acute lymphocytic Leukemia (Segmented) | Swin Transformer-Base + ResNet-50 (Hybrid Model) | train: 70%; test: 15%; val: 15% | 99.65 |

The superior performance can be attributed to the complementary integration of ResNet50's residual convolutional feature extraction with the Swin Transformer's hierarchical self-attention mechanism, enabling effective learning of both fine-grained local textures and long-range contextual dependencies. Notably, the model demonstrates exceptional robustness and generalization capability across heterogeneous datasets, surpassing both traditional ML approaches and advanced CNN-based frameworks reported in recent studies. Furthermore, the observed performance gains on segmented leukemia data highlight the model's ability to leverage region-focused representations without compromising stability. Overall, the comparative results in Table 7 validate that the proposed hybrid framework establishes a new benchmark in multi-cancer histopathological image classification, offering a scalable and clinically viable solution suitable for real-world diagnostic applications.

## 7. Conclusion

We proposed a comprehensive multi-cancer histopathological image classification framework utilizing an integrated architecture that combines robust Swin Transformer with transfer learning techniques. We address critical challenges in cancer diagnosis, particularly in resource-constrained healthcare settings where accurate and timely identification is urgently needed. The proposed model demonstrates exceptional performance, achieving over 99% accuracy with near-perfect precision, recall, and F1-scores across multiple cancer types. The close alignment between training and validation learning curves indicates stable generalization without overfitting, confirming the robustness of our approach. To enhance clinical interpretability and transparency, we employed LIME and occlusion sensitivity analysis across all datasets. These explainability methods demonstrate that the model accurately identifies clinically relevant features, including abnormal cellular clusters, tumor boundaries, and tissue density variations—the same morphological patterns utilized by expert pathologists. This interpretability is essential for building clinical trust and facilitating effective human-AI collaboration in cancer diagnosis workflows. Benchmarking against recent state-of-the-art methods confirms that our proposed framework consistently outperforms existing approaches for both single-cancer and multi-cancer classification tasks, establishing a new performance benchmark in automated histopathological image analysis.

**Ethical Declaration**

This study utilized multi-cancer datasets which were collected from another source. Therefore, no ethical declaration is required for this research.

**Funding Statement**

This research received no specific grant from any funding agency in the public, commercial, or not-for-profit sectors.

**Author Contribution**

**Muazzem Hussain Khan, Tasdid Hasnain, and Ruhul Amin:** Conceptualization, Methodology, Software, Data curation, and Writing – original draft. **Muazzem Hussain Khan, Tasdid Hasnain and Md. Jamil Khan:** Validation, Formal analysis, Visualization, and Writing – original draft. **Ruhul Amin, Md.** Supervision, Conceptualization and Writing – review & editing. **Shamim Reza, and Md. Al Mehedi Hasan:** Conceptualization, review & editing. **Md Ashad Alam:** Supervision, Resources, Conceptualization, and review & editing. All authors have read and agreed to the published version of the manuscript.